\documentclass[a4paper]{amsart}

\usepackage{setspace}
\usepackage[dvips]{graphicx}
\usepackage[]{amsfonts}
\usepackage[]{cite}
\usepackage{amsmath}

\newtheorem{isodefn}{Definition}
\newtheorem{isomorphism}{Theorem}
\newtheorem{sharpnfl}[isomorphism]{Theorem}
\newtheorem{permviol}[isomorphism]{Theorem}
\newtheorem{revisiting_2}[isomorphism]{Theorem}

\newtheorem{performance}{Corollary}

\newtheorem{sensible}[isodefn]{Definition}
\newtheorem{revisiting}{Proposition}

\newtheorem{random}[performance]{Corollary}
\newtheorem{noinfo}[performance]{Corollary}

\newcommand{\CUP}{c.u.p.}
\newcommand{\Hyper}[1]{#1^{#1}}

\begin{document}

\title{Beyond No Free Lunch: Realistic Algorithms for Arbitrary Problem Classes}
\author{James A. R. Marshall$^*$, Thomas G. Hinton}
\address{Department of Computer Science\\
University of Bristol\\
Bristol BS8 1UB\\
United Kingdom\\
$^*$email: James.Marshall@bristol.ac.uk}

\maketitle
\section*{Keywords}

combinatorial problems, search, neutrality, permutation closure, randomised
algorithms, No Free Lunch

\section{Introduction}
The various No Free Lunch theorems are important theoretical results, indicating
that no `black-box' problem solver can be expected to achieve better than random
performance without any information about the problem it is to be applied
to. Such results have been proved for off training-set generalisation in
supervised learning
\cite{Wolpert:1996}, and for search and optimisation \cite{Radcliffe:1995, Wolpert:1997}. It is known that a crucial assumptions of the No Free Lunch theorems is
 that a (block) uniform distribution holds over the set of problems, or objective functions, under consideration and that this set be closed-under-permuation (\CUP). The \CUP~requirement means that any permutation of any objective function in the problem set under consideration, through changing the mapping between elements of the search space and objective values, results in another objective function that is also a member of that problem set. However, it has been shown that realistic problem sets are highly unlikely to satisfy the \CUP~assumption \cite{Igel:2005}. As a result, a large field of increasingly sophisticated work has developed seeking to prove that `Almost No Free Lunch' results that are nearly as strong as the original No Free Lunch results, typically based on information theoretic considerations, still apply in realistic problem scenarios (\textit{e.g.} \cite{Droste:2002, English:1999, Whitley:2008}).

In this paper we take a simplifying step back in an attempt to cut the increasingly complicated `Gordian knot' that (Almost) No Free Lunch research has come to represent. Our approach is simple and intuitive, yet allows general results easily to be arrived at.
We begin by examining the implications of the search version
of the No Free Lunch theorem for real-world search algorithms, showing that
revisiting algorithms also break the permutation closure condition required for the
No Free Lunch theorem to hold. Thus allowing realistic, revisiting algorithms means there can be some best algorithm; the pertinent question then is whether we can identify this best algorithm, and it turns out that we can indeed.
To answer this question we go on to present a novel analysis of
search algorithms as stochastic processes, enabling us to present a No Free
Lunch-like result that holds for arbitrary sets of problems, and for realistic
algorithms that do not avoid revisiting points in the search space. Specifically we show that random enumeration has the best expected performance of all search algorithms applied to optimisation problems, for any distribution over any problem set. The implication of this is that empirical demonstration of superior search performance (relative to enumeration) for some algorithm on some problem set still predicts inferior performance on some second problem set if we know nothing about its relationship with the first. We thus `cut the Gordian knot' by
simplifying the assumptions underlying the No Free Lunch theorem, and show why
violations of its assumptions are unimportant for real-world search algorithms
and problems.

\subsection{The Sharpened No Free Lunch Theorem}
We begin by informally summarising the Sharpened No Free Lunch theorem. No Free
Lunch arguments typically consider a search space $\mathcal{X}$, a set of
possible objective values $\mathcal{Y}$, and objective functions of the form $f:
\mathcal{X} \mapsto \mathcal{Y}$. The set of all possible objective functions is
then denoted $\mathcal{F}=\mathcal{Y}^\mathcal{X}$. The original No Free Lunch
theorems prove that, across all possible objective functions (or problems)
$\mathcal{F}$, all non-revisiting algorithms have equivalent performance under
an arbitrary performance measure \cite{Wolpert:1997, Radcliffe:1995}. Algorithms
are defined here as pseudo-random processes choosing previously unvisited points
to visit based on the quality of prior visited points. Subsequently it was shown
that the No Free Lunch theorem only holds if the set of objective functions
under consideration is closed under permutation (\CUP).

The resulting theorem \cite{Schumacher:2001, Igel:2003a} can be paraphrased as:

\vspace{1.5ex}

\begin{sharpnfl}[Sharpened No Free Lunch Theorem] \label{sharpnfl} All
  non-revisiting algorithms have equivalent performance over a set of objective
  functions $F$, under some arbitrary performance measure, iff $F$ is closed
  under permutation.
\end{sharpnfl}

\vspace{1.5ex}

Theorem \ref{sharpnfl} has subsequently been extended to also hold for
`block-uniform' distributions over subsets that are \CUP \cite{Igel:2005} \cite{Rowe:2009}. This
condition for the No Free Lunch theorem to hold is satisfied by the original
requirement in \cite{Wolpert:1997} of a uniform distribution over $\mathcal{F}$,
because the uniform distribution is a special case of a block-uniform
distribution. Subsequently it has been shown that permutation closure is very
unlikely to be satisfied in realistic scenarios \cite{Igel:2005}, leading many
researchers to conclude that No Free Lunch results have little consequence for
practical applications of search algorithms.

\section{Revisiting Algorithms}
We now examine one of the fundamental assumptions in the proof of the No Free
Lunch result, that the algorithms used are non-revisiting in the
solution space. That is, the algorithms visit every point in the
solution space exactly once. It has been remarked previously that
such algorithms are impractical due to the time and space cost of storing and
querying the set of visited points \cite{Radcliffe:1995}, although this is not
strictly correct. It has also been suggested that revisiting algorithms, in the
form of algorithms having a degree of redundancy in their solution encoding,
have much to recommend them \cite{Igel:2003b}.

If we relax the assumption of non-revisiting algorithms, we can demonstrate the
following result.

\vspace{1.5ex}

\begin{permviol}[Revisiting Breaks Permutation Closure] A revisiting search over
  a given search space under a given \CUP~set of objective functions can be formally expressed as a non-revisiting search over some larger search space under a
  set of objective functions that is not \CUP, for any set containing only
  objective functions mapping to more than one element.
    \label{permviol}
\end{permviol}
\vspace{1.5ex}
\emph{Proof: } Searching under a \CUP~set of objective functions $F$ with an
algorithm $A$ which revisits some points $x_i\in\mathcal{X}$ is equivalent to a
non-revisiting search of a new set of extended objective functions $F'$ on a
larger space $\mathcal{X}'$, which contains $i$ additional points $x_i'$ under
the constraint that $\forall i : f(x_i) = f(x_i')$. 

For $F'$ to be closed-under-permutation it must definitely be larger than $F$,
as it contains functions with the same codomain (because the search is presumed
to be eventually exhaustive) on the larger domain $\mathcal{X}'$, and increasing
the size of the domain necessarily increases the number of possible
permutations. However, by construction $|F| = |F'|$, and so $F'$ cannot be
closed-under-permutation. $\Box$ \vspace{1.5ex}

Note that this theorem holds regardless of the degree of revisiting, so algorithms which
revisit all points in the search space with equal frequency also effectively
break permutation closure of the set of objective functions.

We discuss the implications of this theorem for previous No Free Lunch
results on encoding redundancy \cite{Igel:2003b} in section \ref{sub:encoding-redundancy-neutrality-and-revisiting}. For now we
briefly note the following obvious but simplifying result:
\vspace{1.5ex}

\begin{random}[Free Lunches]\label{freelunches} All violations of the No Free
  Lunch theorems can be expressed as non-block-uniform
  distributions over problem subsets that are closed under permutation.
\end{random} 

\vspace{1.5ex}

\emph{Proof:} For uniform distributions over \CUP~sets this follows
directly from theorem \ref{permviol} and the observation that non-\CUP~sets are
special cases of non-uniform distributions over \CUP~sets. The extension to
block-uniform distributions is by applying theorem \ref{permviol} to each
\CUP~subset. $\Box$

\vspace{1.5ex}

It is interesting to note that while violation of permutation closure leads to search algorithms potentially having different performance, this does not tell us to what extent algorithms might differ in performance. If induced performance differences are small, then searching for a superior algorithm for some problem set could be like looking for a `needle in a haystack'. Here we present a first approach to quantifying the extent to which revisiting algorithms differ in performance, according to the amount of revisiting they allow. 
That violation of permutation closure leads to algorithms potentially having
different performance has been known since \cite{Schumacher:2001, Igel:2003a}.
However, the strength of the Sharpened NFL theorem makes its contrapositive
correspondingly weak; if $F$ is not \CUP, then there exists at least one
performance measure under which at least one pair of algorithms have differing
performance on $F$.  This leads to the relevant question of how likely two
randomly selected algorithms are to have different performance for a given
non-c.u.p.~problem set, which has not previously been addressed. If most
algorithms have identical performance on a given problem set, then looking for an
algorithm with superior performance may be like looking for a needle in a
haystack. If, however, almost all algorithms have different performance then empirical
comparison of different algorithms becomes sensible. In fact the probability of
two arbitrarily chosen algorithms having a different set of traces over some problem set, and hence potentially different performance under some appropriate performance
measure, grows super-polynomially with increased frequency of revisiting, and so
we can make a stronger claim than the contrapositive of the Sharpened No Free
Lunch theorem provides: when revisiting is allowed, performance
differences can be expected. The proof for this claim follows:

\vspace{1.5ex}
\begin{revisiting_2}
  Let $A$ be an algorithm which performs a revisiting search on a space
  $\mathcal{X}$ and revisits $r$ points before exhausting $\mathcal{X}$. The
  probability that $A$ and a similar algorithm $B$ are indistinguishable in
  performance decreases super-polynomially as the amount of revisiting
  increases, except where all elements of the search space are mapped to the same objective value.
\end{revisiting_2}
\vspace{1.5ex}

\emph{Proof: } Let $F'^*$ be the permutation closure of $F'$, and $F={f}^*$
where ${f}^*$ is the permutation orbit of function $f$. Let
$\mathcal{C}\subseteq\mathcal{Y}$ be the codomain of $f$, and let $\lambda_i$ be
the number of points in $\mathcal{X}$ which $f$ maps to $i\in\mathcal{C}$
($\lambda$ is the \emph{histogram} of $f$ \cite{Igel:2005}).  Searching a set of
functions takes each function to a trace, so the size of an algorithm's trace
set is $|F'| = |F| = \frac{|\mathcal{X}|!}{\prod_{i\in\mathcal{C}}
  \lambda_i!}$. Since $F'$ is not \CUP~this set of traces will be a small
fraction of the traces for $F'^*$; we must make some assumptions about the
behaviour of our algorithm to find $|F'^*|$. Consider the case where the
algorithm revisits points assigned to the most common objective value,
$j\in\mathcal{C}$, $r$ times during the search -- the algorithm may do this by,
for example, enumerating $\mathcal{X}$ until it knows $j$ and
then revisiting a point mapped to it $r$ times -- this simplistic revisiting algorithm is easy to analyse, and upper bounds the probability of distinguishability by ensuring that every function in $F'$ has the same histogram and minimising the size of its permutation closure.   Knowing that all the functions in
$F'$ have the same histogram, constructed by distributing the $r$
revisits amongst the $i$ bins of $\lambda$, the upper bound follows from the observation
that $\prod_i \lambda'_i !$ is maximised by allocating all $r$
revisits to the largest bin in $\lambda$ (this claim can be shown by a
simple induction).

Since every function in $F'$ has the same histogram, in which $\forall x \neq j :
\lambda'_x = \lambda_x$ and $\lambda'_j = \lambda_j+r$, $|F'^*| =
\frac{(|\mathcal{X}| + r)!}{\prod_{i\in\mathcal{C}}(\lambda_i + [i=j] \cdot
  r)!}$. The ratio $p = \frac{|F'|}{|F'^*|}$ is then the fraction of the set of
all possible traces which a given algorithm will produce. Observing that a
uniformly randomly chosen algorithm searching a given objective function gives a
uniform distribution over the traces for that function, we see that for an
algorithm $A$ searching $F'$ an arbitrarily chosen algorithm $B$ shares a trace
with $A$ with probability $p$, and so $p$ bounds from above the probability that
$A$ and a random $B$ are indistinguishable.

Some manipulation gives that $p = \frac{|\mathcal{X}|!}{(|\mathcal{X}|+r)!} \cdot
\frac{(\lambda_j+r)!}{\lambda_j!}$, from which we can show $\ln p = \sum_{i=1}^{r}
\ln\frac{\lambda_j+i}{|\mathcal{X}|+i}$. We can upper-bound this with an integral, which
gives us

\[
p(r) = \frac{|F'|}{|F'^*|} <
\frac{\Hyper{(r-1+\lambda_j)}\times\Hyper{|\mathcal{X}|}}{\Hyper{\lambda_j}\times\Hyper{(r-1+|\mathcal{X}|)}},
\]

from which we can see that $1/p(r)$ is at least superpolynomial in $r$
(excepting the trivial case where $\lambda_j = |X|$). $\Box$ 
\vspace{1.5ex}

Thus the probability that two algorithms selected uniformly at random are
distinguishable increases very quickly as the extent to which the algorithms
revisit increases. The expected number of points needing to be evaluated before a difference is found, and hence the computational complexity of detecting differences in algorithms' performance, is an interesting question that is outside the scope of the current paper.
Furthermore, this approach will still not enable us to reason about \emph{relative} performance of algorithms.

Returning to the main argument of the paper, we will next propose an alternative way of reasoning about revisiting algorithms and
their expected performance over arbitrary sets of possible objective functions. This approach will indeed allow us to reason about algorithms' relative performance.

\section{Realistic Performance Measures and Random Search Algorithms}

Proving No Free Lunch for arbitrary performance measures is a powerful result,
based on an unrealistic assumption; that search algorithms are described by
exhaustive non-revisiting enumerations of the search space and, equivalently
that the set of objective functions considered be closed under permutation.

For the remainder of this paper we change our viewpoint from regarding search
algorithms as deterministic exhaustive, possibly repeating, searches, to
regarding them as stochastic processes. We also change from considering
arbitrary performance measures, to considering performance measures
representative of the goals of search and optimisation. These two changes in
perspective will enable us to reason about the expected performance of realistic
search algorithms, and make concrete recommendations to the practitioner.

First we define a \textit{sensible} performance measure. As mentioned above,
showing that No Free Lunch does not hold does not enable us to reason usefully
about relative performance of different algorithms. We therefore define a
sensible performance measure that reflects the goal of search and optimisation
algorithms: finding good points in the search space. Concentrating on a
particular performance measure will allow us to make definitive observations
and recommendations on the relative performance of different search
algorithms. This would not be possible if we allowed ourselves arbitrary
performance measures (since an arbitrary performance measure is not
\emph{required} to distinguish between algorithms simply because some
performance measure \emph{could} do so).

\vspace{1.5ex}

\begin{sensible}[Sensible performance measures] \label{sensible} Let $S(A, f,
  n)$ be the set of distinct values of objective function $f$ observed by
  algorithm $A$ after $n$ points have been evaluated. Then a \emph{sensible
    performance measure} $M$, which we wish to maximise, defines the performance
  of $A$ on $f$ after $n$ points using only $S(A, f, n)$, with the additional
  constraint that $S_1 \subseteq S_2 \implies M(S_1) \leq M(S_2)$
\end{sensible}

\vspace{1.5ex}

Such measures capture the basic criterion against which search algorithms are
assessed: how good are the solutions they generate? At first, it may appear that
this performance measure discards the other important criterion of search
algorithm performance: how long does it take to generate good solutions?
However, as we shall see time is implicit in this performance measure, as we
shall make comparisons of searches having equal length.

We now make the following No Free Lunch-like statement for revisiting algorithms.

\vspace{1.5ex}

\begin{revisiting}[Maximising Sensible Performance] \label{enumeration} For any
  distribution over any set of objective functions $F$, a randomly chosen
  enumeration $A$ can be expected to equal or outperform a randomly chosen
  non-minimally-revisiting algorithm $B$, under any sensible performance measure
  $M$ at any time $t$ in the search.
\end{revisiting}
\vspace{1.5ex}

\emph{Proof: } For any $t \geq |\mathcal{X}|$ it is clear that a minimally
revisiting algorithm has maximal performance as the full codomain of the
objective will have been observed and, from definition \ref{sensible}, $M$ is required to depend only on the part
of the codomain which has been seen.

When $t < |\mathcal{X}|$, $A$ samples $t$ points from $\mathcal{X}$ without
replacement, whereas $B$ samples $t$ points from $\mathcal{X}$ with some
non-zero probability of replacement (as a consequence of being
non-minimally-revisiting). Consequently the expected number of distinct points
sampled by $A$ is greater than the expected number sampled by $B$. Similarly the
expected size of the set of objective values observed by $A$ exceeds (or equals,
for functions whose codomain is of size 1) that of the set observed by $B$. Call
these two sets of objective values $S_A$ and $S_B$.

$S_A$ and $S_B$ are subsets drawn randomly from the codomain of the objective,
and the only effect $B$ can have on the distribution of values drawn is to
reduce its range, so that $|S_B| \leq |S_A|$. For a randomly chosen set $S$ of
objective values, $E(M(S))$ increases with $|S|$ (this is clear from the
definition of a sensible measure), and so $E(M(S_A)) > E(M(S_B))$. This holds for any individual objective function, the extension to arbitrary distributions over arbitrary sets being by linearity of expectation. $\Box$
\vspace{1.5ex}


It is sometimes remarked that No Free Lunch approaches typically ignore the time and space complexity of the algorithms, therefore it is interesting at this point to note that enumeration can typically be implemented with excellent time and space complexity; $O(log |\mathcal{X}|)$ time for each point queried\footnote{The time taken to exhaust the search space is thus $O(|\mathcal{X}|)$, which also cannot be improved upon.} and $O(\log |\mathcal{X}|)$ space overall, when solutions can be represented as finite strings. No algorithm with lower space complexity can have better performance in the NFL sense (it would necessarily revisit, as it could not have a state for each point in which it would deterministically query that point), and only algorithms with better performance in the NFL sense can optimise faster in wall-clock time. This further indicates enumeration's pre-eminent position as the general search algorithm, and motivates its use as a benchmark by practitioners.

An interesting consequence of proposition \ref{enumeration} is

\vspace{1.5ex}
\begin{random}[Random Search Performance]\label{randomsearch} Random search
  algorithms,
those that ignore the outcome of their search in selecting subsequent search points, can
differ in performance.
\end{random}
\vspace{1.5ex}

\emph{Proof:} This follows directly from proposition
\ref{enumeration} and observing that random search algorithms, as defined above,
include both blind enumeration, and revisiting algorithms.  $\Box$
\vspace{1.5ex}

The definition of random search algorithms as being those that ignore the
objective values generated during the search was proposed by
\cite{Wolpert:1997}. Corollary \ref{randomsearch} is interesting, as the
fundamental prediction of the No Free Lunch result has been stated as being that
``if an algorithm performs better than random search on some class of problems
then it must perform \emph{worse than random search} on the remaining problems''
(emphasis authors') \cite{Wolpert:1997}. Corollary \ref{randomsearch} shows that
the property of `randomness' is actually not of great importance for No Free
Lunch-like statements regarding realistic search algorithms.

Finally, an important result emerges as a direct consequence of proposition \ref{enumeration}:

\vspace{1.5ex}
\begin{noinfo}[Performance Prediction]\label{noinfo} Empirically-demonstrated better-than-enumeration performance of some algorithm on some problem set predicts worse-than-enumeration performance on any other disjoint problem set whose relationship to the first is unknown.\end{noinfo}
\vspace{1.5ex}

\emph{Proof:} For some algorithm A let us denote expected deviation from the expected performance of enumeration on a problem set $\psi$ as $E(\Delta_A(\psi))$. Then, from proposition \ref{enumeration}, $E(\Delta_\textrm{enum}(\psi))=0$ for any $\psi$, and similarly for any non-minimally-revisiting algorithm $A$ whose performance on $\psi$ we know nothing about, $E(\Delta_A(\psi))<0$. If we have an unknown problem set $\phi$ whose relationship with set $\psi$ is unknown then, by definition, $E(\Delta_\textrm{A}(\phi) | \psi) = E(\Delta_\textrm{A}(\phi))$, which must also be negative by proposition \ref{enumeration}.  $\Box$
\vspace{1.5ex}

A related but different result based on considerations of function complexity has previously been presented \cite{Droste:2002}, which demonstrates that for any function where an algorithm performs well there is another function \emph{of similar complexity} on which it performs badly. Our result is both simpler and more generally applicable, in that it makes no reference to the details of the algorithms or the complexity of the functions involved, but weaker in its predictions, in that it predicts no correlation between performances on the two problem sets, rather than a negative relation.

\section{Encoding Redundancy, Neutrality and Revisiting}
\label{sub:encoding-redundancy-neutrality-and-revisiting}
In No Free Lunch contexts, revisiting by an algorithm might occur through the
decision not to keep track of search points visited so far. An additional cause
for revisiting is encoding redundancy. We can represent encoding in a No Free
Lunch context by considering an additional set $\mathcal{W}$ of encoded
solutions, which are decoded into their corresponding points in the search space
by a `growth' function $g: \mathcal{W} \mapsto \mathcal{X}$
\cite{Radcliffe:1995}. If this mapping is not injective, then we have a
redundant encoding. Such redundancy is often found in heuristic search
algorithms, and will turn a non-revisiting algorithm in the space of encoded
solutions into a revisiting algorithm in the true solution space, so that the No
Free Lunch theorem no longer holds. Radcliffe \& Surry may not have noticed
this fact, as their method of proving the No Free Lunch result relies on setting
the search space $\mathcal{W}$ and solution space $\mathcal{X}$ to be the same,
and the growth function $g$ to be the identity mapping. It has previously been
suggested in \cite{Woodward:2003} that a non-uniform (\textit{i.e.}  biased)
encoding redundancy would break the No Free Lunch theorem. Here we have
demonstrated that in fact non-uniformity is not the key requirement for breaking
No Free Lunch, encoding redundancy alone is sufficient.

Previously Igel and Toussaint have presented an analysis of encoding redundancy,
or neutrality, in No Free Lunch situations, arguing that under encoding
redundancy all algorithms still have equal expected performance
\cite{Igel:2003b}. In fact the No Free Lunch theorem does not hold once
redundancy is introduced for, as shown above, redundant and hence revisiting
algorithms break permutation closure of any set of objective functions. Thus
care must be taken in interpreting these results; the assumption of theorem 1 in \cite{Igel:2003b} is violated and hence the subsequent claim that the time to find an optimum averaged over all functions is the same for all algorithms when redundancy is allowed is incorrect. Igel and Toussaint actually
analyse the difficulty of search problems in terms of the proportion of
\emph{distinct} solutions  (members of $\mathcal{X}$) mapped to \emph{the same} global optimum (member of set $\mathcal{Y}$); this is not the same thing as considering multiple encoded solutions (members of $\mathcal{W}$) mapped to the same actual solution (member of set $\mathcal{X}$).
 As their analysis considers only
algorithms that are non-revisiting in the solution space the No Free Lunch
result is still applicable and their analysis valid, however nothing has been
said about the effects of true encoding neutrality on the performance of search
algorithms. When encoding redundancy is correctly defined as a non-injective
mapping between a larger representation space and a smaller solution space, the
stochastic search results presented here become applicable. The observation that
encoding redundancy in a given problem leads to enumeration having the best
expected performance may raise questions about the theoretical basis for the
claimed benefits of neutrality and self-adaptation \cite{Igel:2003b}, which are outside the scope of this paper.

\section{Conclusions}
In this paper we have shown that the requirement for the Sharpened No Free Lunch
theorem to hold, that the problem set under consideration be closed under
permutation, is in fact a necessary and sufficient condition for a version of the theorem, extended to also admit revisiting algorithms, to hold. It has previously been shown that the \CUP\ condition is very unlikely to be satisfied for realistic problem classes. We have shown further that realistic, revisiting algorithms also violate this condition.
That the No Free Lunch theorems' assumptions are typically violated has been used many times as an argument that they can reasonably be ignored by those working with search algorithms, while a branch of research has also grown up showing conditions under which `Almost' No Free Lunch results hold.


If the conditions of No Free Lunch are violated then some algorithm can have best expected performance, but this does not help us to find it, or to say anything about it.
The main contribution of this paper is to propose a statistical analysis of
search algorithms as random processes, demonstrating that algorithms that
minimise the extent to which they revisit points in the search space have higher
expected performance. The result of this is the demonstration that for arbitrary
sets of objective functions, including those which are not closed under
permutation and over which arbitrary probability distributions hold, enumeration
has better expected sensible performance than any arbitrary revisiting
algorithm. Note that arbitrary problem distributions obviously includes
realistic problem distributions, such as the universal distribution
\cite{Solomonoff:1964}. It is also interesting to note that, through use of a
string encoding, an algorithm to perform a blind enumeration of a search space
will typically have excellent algorithmic time and space complexity, in $O(|\mathcal{X}|)$ and $O(\log |\mathcal{X}|)$ respectively.


This paper may also help guide practice. Theorem \ref{noinfo} should be of great interest to those empirically investigating the performance of search algorithms. It does not say that superior performance on one problem set will necessarily result in inferior performance on another, but does say that if we have no information whatsoever about the relationship between two problem sets, performance observed to be better than enumeration on one still predicts performance worse than enumeration on the other. Of course, it is rare that one knows nothing at all about the relationship between problem sets, even if it is simply that at an intuitive level problems from the two sets seem similar in some way. Furthermore as an experimenter collects performance statistics for their algorithms on the unknown problem set, uncertainty over the algorithms' performance on that set will decrease so that they are able to predict performance on previously unseen instances with increasing confidence. Nevertheless, consider the best policy if you were asked to play the following simple game: a third-party gives you an algorithm and a set of performance data for it on some problem set, then tells you that you have a choice of whether to apply the algorithm to one problem from some other problem set, or to apply a random enumerative search. No information whatsoever is given about the new problem set, and the aim of this game is to achieve the best performance on the new unseen problem on your first attempt. Theorem \ref{noinfo} indicates that you should choose enumeration instead of the known algorithm. Obviously one rarely, if ever, wishes to win such a game, and applying a known algorithm to an unknown problem set has another payoff in terms of information gained, but hopefully theorem \ref{noinfo}, the computational simplicity of enumeration, and the game outlined above will all help to convince practitioners that an algorithm's empirical performance on some problem set
should be
directly compared against a deterministically enumerative search started at some
uniformly randomly selected point in the search space or, equivalently, a
uniform random search using sampling without replacement. At present such
practice is far from common.

\section*{Acknowledgments}
We thank D. Wolpert, S. McGregor, R. Clifford, A. Harrow, T. Kovacs and
certain anonymous reviewers for helpful discussions and comments.



\bibliographystyle{elsart-num-sort.bst}
\bibliography{nfl_cec.bib}
%
\end{document}